\documentclass[DIV15,a4paper]{scrartcl}

\usepackage[pdftex]{graphicx}
\graphicspath{{./figures/}}
\DeclareGraphicsExtensions{.pdf,.jpg,.png}
\usepackage{tabularx}

\begin{document}

\title{Bridging Ecology and Cloud: Transposing Ecological Perspective to Enable Better Cloud Autoscaling}
\author{Tao Chen, Rami Bahsoon \\ Centre of Excellence for Research in Computational Intelligence and Applications\\
School of Computer Science, University of Birmingham\\
Edgbaston, Birmingham, UK, B15 2TT\\
\{t.chen, r.bahsoon\}@cs.bham.ac.uk}

%
%
\maketitle

\abstract{ Elastic autoscaling is the fundamental mechanism that enables the cloud-based services to continually evolve themselves\textemdash through changing the related software configurations and hardware resource provisions\textemdash under time-varying workloads. However, given the increasingly complex dynamic, uncertainty and trade-offs related to the runtime QoS and cost/energy of services, cloud autoscaling system is becoming one of the most complex artifacts constructed by human and thus its effectiveness is difficult to be preserved. In this article, we present novel ideas for facilitating cloud autoscaling. Our hypothesis is that \textit{cloud ecosystem}, represented by a collection of cloud-based services, bears many similarities with the natural ecosystem. As such, we intend to investigate how ecological view can be adopted to better explain how the cloud-based services evolve, and to explore what are the key factors that drive stable and sustainable cloud-based services. To achieve this goal, we aim to transpose ecological principles, theories and models into cloud autoscaling analogues and spontaneously improve long-term stability and sustainability of cloud ecosystem.}

\section {Introduction}
In cloud environment, the quality of service (QoS) and cost/energy objectives for cloud-based services can be tuned by accessing software configurations (e.g., the number of service threads, size of connection pool and session lifetime) and hardware resources (e.g., the CPU, memory and Virtual Machine) that are shared, leased, and priced as utilities. Such feature is fundamentally facilitated through autoscaling: an automatic and elastic process, typically running on a Physical Machine (PM), that adapts software configurations and hardware resources provisioning on-demand according to the changing environmental conditions (e.g., the workload).

What make the cloud autoscaling challenging are the dynamic, uncertainty and possible trade-offs on objectives (i.e., QoS and cost/energy objectives) exhibited in the process. In cloud, dynamics and uncertainty are arise from the unpredictable environmental conditions and QoS interference\textemdash a scenario where the competing demands of some of the services can interfere with the QoS of others, providing that there are many cloud-based services run on a shared infrastructure. Trade-offs are associated with runtime autoscaling decisions as to the appropriate amount of scaling applied to software configurations and hardware resources. These trade-offs can be, for example, whether to choose throughput over cost or which competing cloud-based services to focus on.

To address those challenges, prior efforts in cloud autoscaling have relied on rules-based control \cite{scale-rule-based}, control theoretic mechanisms \cite{TR-10-full-version} and computational intelligence techniques \cite{Chen:2014}, with particular focus on improving the optimality, elasticity and scalability of the autoscaling system. However, the runtime stability and sustainability in the cloud, which are mainly concerned with the long-term benefit for both cloud consumers and provider, have not been explicitly tackled. Here, \textbf{\emph{sustainability}} refer to the ability of cloud to endure the stress caused by dynamic and uncertain events, e.g., workload and QoS interference, with an aim to continually optimize QoS attributes of all cloud-based services while minimizing their costs and energy consumptions. The longer the time that no violation of Service Level Agreement (SLA) and budget requirements occur, the better the \textbf{\emph{stability}} is. Following the intuition that \textit{computer systems can be better understood, controlled, and developed when viewed from the perspective of living systems} \cite{Forrest02computationin}, we argue that the perspective of ecology and natural ecosystem \footnote{we will discuss this in details in Section \ref{sec:ne}} is new, yet neat view for computer science researchers to design novel autoscaling system in the cloud. In particular, understanding stability and sustainability of natural ecosystem, as well as how we can better manage them spontaneously, have been the core research theme for ecologist. These will therefore provide many useful insights for researches in cloud autoscaling.

In this article, we explore on the potential benefit of using ecological view when designing autoscaling system in the cloud. By ecological view, we refer to render the cloud environment as a natural ecosystem and to design autoscaling system in the cloud derived/inspired from ecological techniques. We then propose a sensible translation of ecological principles, theories and models into cloud autoscaling analogues. Particularly, we have explored the biotic characterises of cloud-based services and the underlying cloud primitives w.r.t. the principles of living organisms/species, nonliving components, habitats, biodiversity, disturbances, species competition, trophic web and natural evolution.  We propose an ecology-inspired self-aware architectural pattern extending on the self-aware patterns from our prior work \cite{chen2014handbook}. The new pattern explicitly caters for the key levels of biotic information, which are also systematically linked to the original principles of self-awareness. Finally, we highlight the challenges and opportunities for future investigations of ecology-inspired autoscaling in the cloud.

\section {Motivation}
  
Elastic autoscaling in the cloud has been an increasingly important research topic since the emergence of cloud computing paradigm. Efforts have been spent to deal with the dynamics, uncertainty and trade-offs exhibited in the autoscaling process \cite{scale-rule-based}\cite{TR-10-full-version}\cite{Chen:2014}. Nevertheless, how autoscaling can improve the stability and sustainability of the cloud as a whole has not been explicitly studied in prior work. Undoubtedly, stability and sustainability are among the most desirable attributes of cloud computing. Table \ref{table:compare} illustrates the benefits of explicitly considering stability and sustainability when autoscaling in the cloud.

Natural ecosystems are considered to be robust, efficient and scalable systems that are capable to cope with dynamics and uncertainty, possessing several properties that may be useful in cloud autoscaling systems. These properties include self-awareness, self-adaptivity, and the ability to provide solutions for complex scenarios \cite{ecosystem}, e.g, resolving trade-offs. Many of these properties can be understood via well-known ecological models \cite{hubbell2001unified}, which provide a theoretical basis for the occurrence of self-awareness and self-adaptivity, resulting from the interactions among the individuals and their environment, leading to complex and emergent behaviors \cite{ecosystem} (e.g., evolution driven by natural selection).

Among others, stability and sustainability are the most desirable attributes in natural ecosystem and they have been studied by the ecologists for decades. We advocate that the well-established ecological principles, theories and models can provide rich source of inspiration to spontaneously improve the stability and sustainability of the cloud as a whole. This will allow all the cloud-based services to stay robust and generate minimal overhead when optimizing their objectives and complying their SLA/budget requirements. However, this begs the question: \emph{how to systematically incorporate the natural ecosystem and cloud autoscaling?}

 \begin{table}[t!]
  \caption{Comparing Autoscaling in the Cloud With/Without Tackling Stability and Sustainability.}
\label{table:compare} 
\begin{center}
 \begin{tabularx}{\linewidth}{|p{5.5cm}|X|}

 \hline \hline
\textbf{With Stability and Sustainability} &
\textbf{Without Stability and Sustainability} \\  
\hline\hline
Improve QoS, cost and energy from a long-term perspective. &
Improve QoS, cost and energy, but might only be effective in short-terms. \\  \hline 
Resilient to extreme cases, e.g., sudden and spiked workloads. &
Vulnerable to extreme cases, e.g., sudden and spiked workloads. \\  \hline
Aim for less scalings and smaller overhead. &
Easy to result in unnecessary scalings and larger overhead. \\  
 \hline \hline
\end{tabularx}
\end{center}
\end{table}

\section {Natural Ecosystem}
\label{sec:ne}
From the ecological perspective, the term \emph{ecosystem} refers to a natural community where all the living organisms (e.g., plants and animals) and nonliving components (e.g., air, light and water) exhibit dynamic, and uncertain interactions with each others and the environment, emerging as a system \cite{pickett1987ecology}.

The dynamics in an ecosystem is represented by trophic web\textemdash an interaction network that models the consumer-resource relationship, for examples, predator-prey or organism-resources. The trophic web often consist of different trophic levels, each of which represents a family of functionally consistent species. The consumer-resource relationship takes place between different trophic levels. The foundation of a trophic web are the autotrophic species (e.g., most plants) who can produce complex organic compounds from nonliving components. In contrast, higher levels in the trophic web are heterotrophic species (e.g., the animals) who cannot fix carbon and uses organic carbon for growth. There is a special kind of mixotrophic species that use a mix of different sources of energy and carbon, e.g., the venus flytrap and oriental hornet.


An ecosystem might face with disturbances causing by either natural or human induced stress, e.g., tornado and deforestation are natural and human induced stress respectively. Sustainability is often refer to the endurance of ecosystem in the presence of disturbances. Better sustainability of an ecosystem implies better stability\textemdash an ecosystem is said stable if, when the disturbances occur, it is not affected or it is able to quickly resume back to its prior stable state after disturbances.

According to the well-recognized insurance hypothesis \cite{yachi1999biodiversity}, ecologists have acknowledged that better stability and sustainability of an ecosystem can be achieved on higher biodiversity. This is because an ecosystem with large diversity of species will be able to respond to the disturbances in different ways, and thus it is more likely to resume to the previous stable state as some species can compensate for those that disappear.

Biodiversity is a corollary of evolution, which describes the ability of a species in survive and reproduce. This is attributed to the fact that a species contains different individuals whose concrete characteristics are vary, e.g., their genetic code and habits. Driven by the natural selection during evolution, new species might be created and the incapable ones might disappear. Evolution might lead to the change of ecological niche, which represents how the species responds to the changes of resources and competitors. Changing the niche could also implies conversion of the habitat (e.g., land or sea) that are used by the species, e.g., the ancestors of whales was living on land. According to the competitive exclusion principle \cite{hardin1960competitive}, evolution is one of the results of competition, in which the benefit of one species can be lowered by the presence of the others\textemdash this might be due to the limited supply of certain resources. A special case, namely co-evolution, occurs if one species changes in response to the changes in others.

\section {Transposing Ecological Principles, Theories and Models to Cloud Ecosystem}

Existing ecology researches have provided many insights on how we can better preserve the natural ecosystem, particularly with an aim to improve its stability and sustainability. Our hope is to learn and investigate how these insights can be used to derive better elastic autoscaling in the cloud. In fact, as we will show below, the collection of cloud-based services operates in a way that has many similarities to the natural ecosystem, and therefore emerging as a \textit{cloud ecosystem}.

\begin{figure}[t!]
\centering
\includegraphics[width=0.6\linewidth]{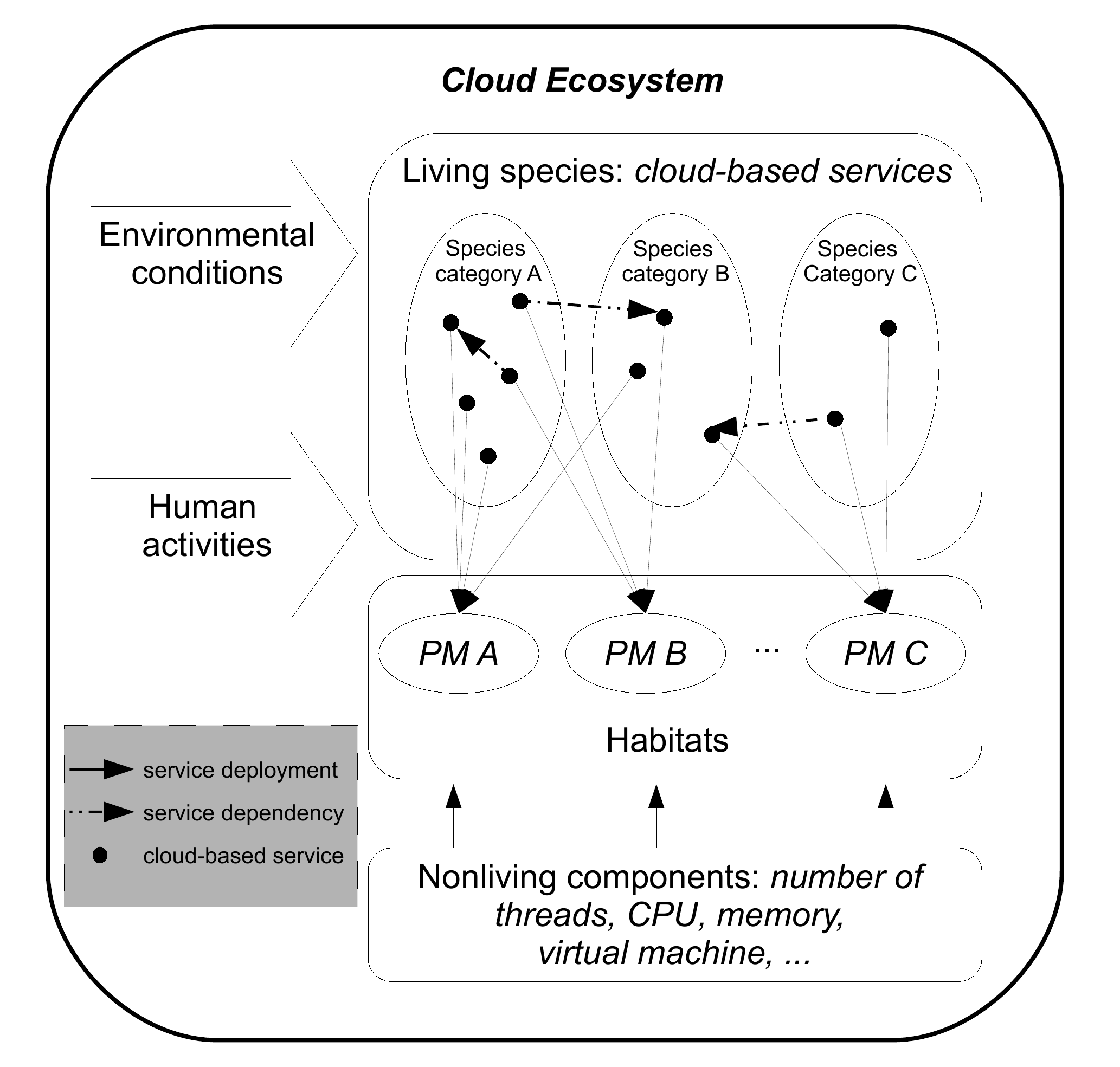}
\caption{The architecture of cloud ecosystem.}
\label{fig:arch}
\end{figure}

\textbf{\emph{Ecosystem}}\textemdash The 3-layered architecture of cloud ecosystem is illustrated in Figure \ref{fig:arch}. At the top layer, the cloud-based services are regarded as living organisms, which are categorised as species within the cloud ecosystem. These species can be fundamentally classified based on the nature of the services, e.g., its functional dependency to other services and whether it is a commercial web service or a scientific service etc. This is similar to the general classification of species in the natural ecosystem, e.g., plants, animals and micro organisms. Such classification might be useful for us to study the characteristics of the species in cloud ecosystem. At the intermediate layer, one or more species might co-exist in a habitat, i.e., a PM that encapsulates the necessary nonliving components. Here, different PMs might be heterogeneous, leading to various forms of habitat. In such context, the nonliving components can be the fine-grained and reusable software configurations (e.g., thread of service) and hardware resources (e.g., CPU and memory), as shown in the bottom layer. Particularly, the software configurations are counterparts of infinite natural resources (e.g., light and air) while the hardware resources are the components that subject to limited supply, e.g., soil and water in the natural ecosystem. Externally, the cloud ecosystem would be affected by disturbances, including both environmental conditions and human activities: the former refers to the factors that are controlled by neither the cloud provider nor service owners, e.g., the workload and the size of incoming jobs. The later, on the other hand, represents the activities that would influence cloud-based services, as conducted by the service owners or cloud provider.

Similar to the natural ecosystem, the cloud ecosystem reacts to the emergent disturbances and preserves stability in the same pattern: When disturbances occur (e.g., sudden changes in workload), the organisms (cloud-based services) would have to amend their demand on nonliving components (e.g., CPU, memory) or demand on the other species of organisms (when the services have functional dependency), in order to survive in the cloud ecosystem. In certain cases, organisms, or even the entire specie, would need to change the habitat (VM migration), creating the chances of multiple species on the same habitats. All these facts imply evolutions that change the biodiversity, e.g., the demand of services, their underlying VM and neighouring services are changed. In the following, we will explain the mapping between ecological principles and cloud autoscaling system in details.

\begin{figure}[t!]
\centering
\includegraphics[width=0.6\linewidth]{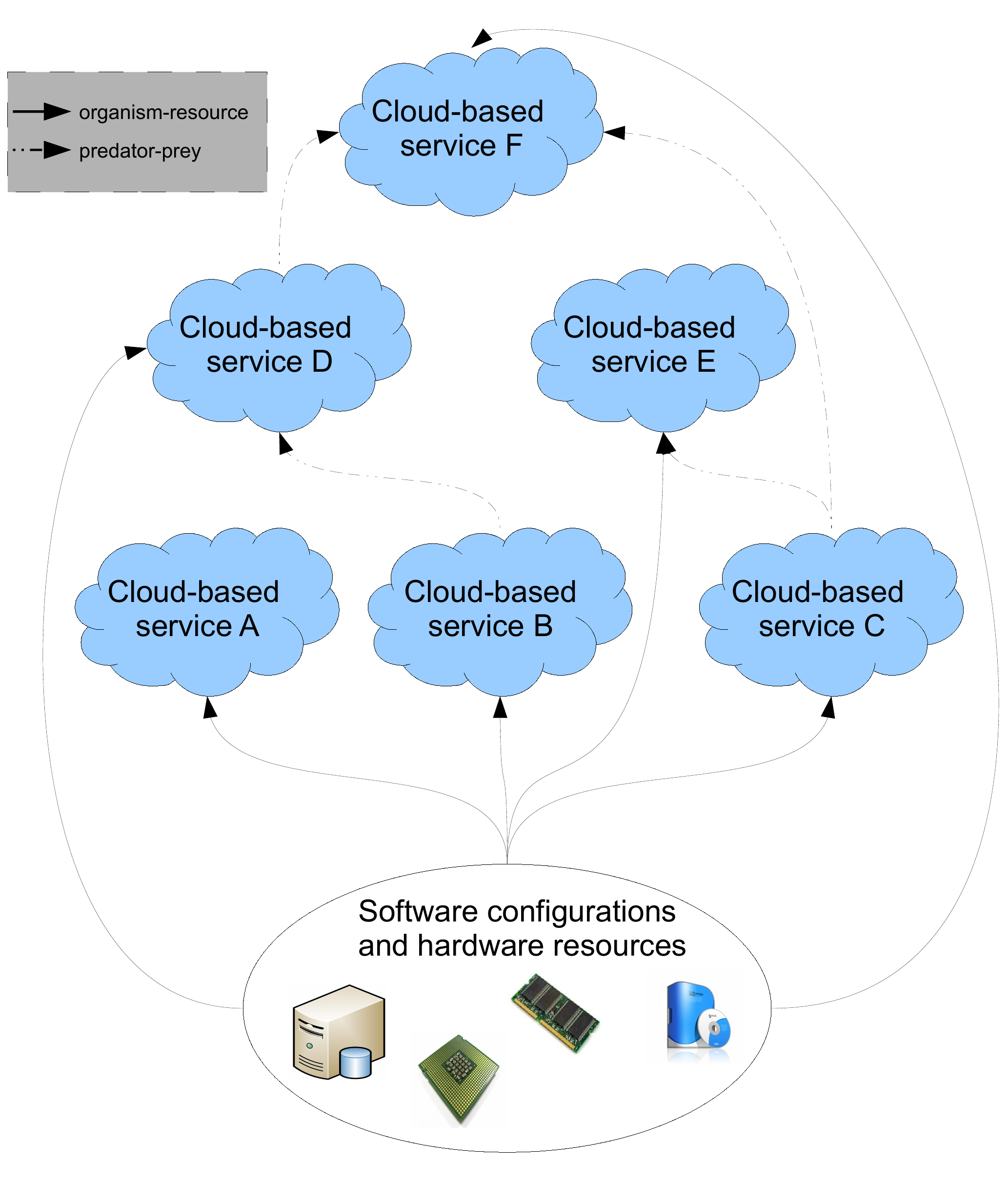}
\caption{The trophic web of cloud ecosystem.}
\label{fig:cloud-food-web}
\end{figure}

\textbf{\emph{Trophic web}}\textemdash The notion of trophic web can also find matched principle in a cloud ecosystem. As we can see from Figure \ref{fig:cloud-food-web}, the functional dependency between cloud-based services can be modeled as predator-prey relationship, e.g., one or more business services (predator) can dependent on a database service (prey). In the natural ecosystem, the more of species A (predator) usually means the less of species B (prey), until at some point in time this A recedes due to a lack of B. In contrast,  the predator-prey relationship the cloud ecosystem would not change the quantity of  services, but could affect their ability in serving requests and jobs. For example, when the number of predator services increases, their prey services can gradually reach its limit in handling requests, which would be equivalent to "die out". Consequently, the predator services would recede due to the "die out" of their prey, unless they can evolve themselves to seek alternatives. On the other hand, the correlation between cloud-based services and the software/hardware resources is clearly a organism-resources relationship. It is obvious to see that the species in cloud ecosystem can only be either autotrophic or mixotrophic, because on one hand, they directly consume the nonliving components. On the other hand, they may be rely on the organic carbon from the other species.

\textbf{\emph{Disturbances}}\textemdash Similar to natural ecosystems, cloud ecosystem exhibits various forms of disturbances caused by the environmental conditions. These disturbances can be natural induced stress, e.g., the changes in co-located services and co-hosted Virtual Machines (VMs), changes of software configurations and hardware resources provisioning, and VMs migration/replication. Human, i.e., the service owners and cloud providers, might also create stress by changing the SLA and budgets requirements, deploying/removing cloud-based services or amending the prices for renting software and hardware resources.

\textbf{\emph{Biodiversity and evolution}}\textemdash In cloud ecosystem, the species, or cloud-based services, differ depending on their QoS sensitivity to different software configurations and hardware resources, as well as their SLA and budget requirements. In addition, there can be a large diversity for the cloud-based services running on the same PM, or habitat. These are clear evidences of biodiversity in the cloud ecosystem. Through elastic autoscaling, evolution in cloud ecosystem refers to change the ability of cloud-based services in accessing the software configurations and hardware resources, and possibly the deployment of services. This fact allows such evolution to directly influence the biodiversity in cloud ecosystem. Precisely, evolution can be regarded in two levels: (i) at micro evolution level, cloud-based services can continually evolve by changing their software configurations and hardware resource provisioning based on their demand. (ii) At macro evolution level, adding/removing cloud-based services and VM migration/replication can cause changes to their habitats.

\textbf{\emph{Species competition}}\textemdash QoS interference is a good example of species competition in cloud ecosystem, as the cloud-based services running on the same habitat are competing for resources when the supply is limited. Consequently, they need to either evolve themselves (through autoscaling) or die out (crash or being removed by the owners due to severe SLA violations). This phenomena in cloud also play as a counterpart to the co-evolution principle in natural ecosystem: to mitigate QoS interference, two or more cloud-based services might need to evolve in response to each others.

\section{Ecology-Inspired Self-Aware Pattern}

We now describe our preliminary research outcome that extends the self-aware patterns to incorporate ecological principles and biotic information for cloud autoscaling. As systematically documented in our handbook \cite{chen2014handbook}, self-aware patterns describe sets of capabilities (i.e., levels of awareness) to acquire knowledge for a \emph{node} which, in the context of self-aware computing systems, can refer to a process, machine or any conceptual part of a software system being managed. Similar to the original self-aware patterns, the ability to acquire knowledge also play a crucial role in ecology-inspired patterns, but it particularly focuses on acquiring different levels of biotic information and knowledge. Specifically, when describing ecology-inspired pattern in cloud autoscaling, a \emph{node} refers to an autoscaling process, which maintains biotic information for and manages a group of species (i.e., cloud-based services) separated by their categories and/or habitats (i.e., PMs). As mentioned in previous sections, different groups of species and the nonliving components form the cloud ecosystem.


A possible ecology-inspired pattern and the corresponding self-aware capabilities are shown in Figure \ref{fig:pattern}. The key capabilities to acquire biotic information in the pattern can be discussed as the following:

\begin{figure}[t!]
\centering
\includegraphics[width=0.6\linewidth]{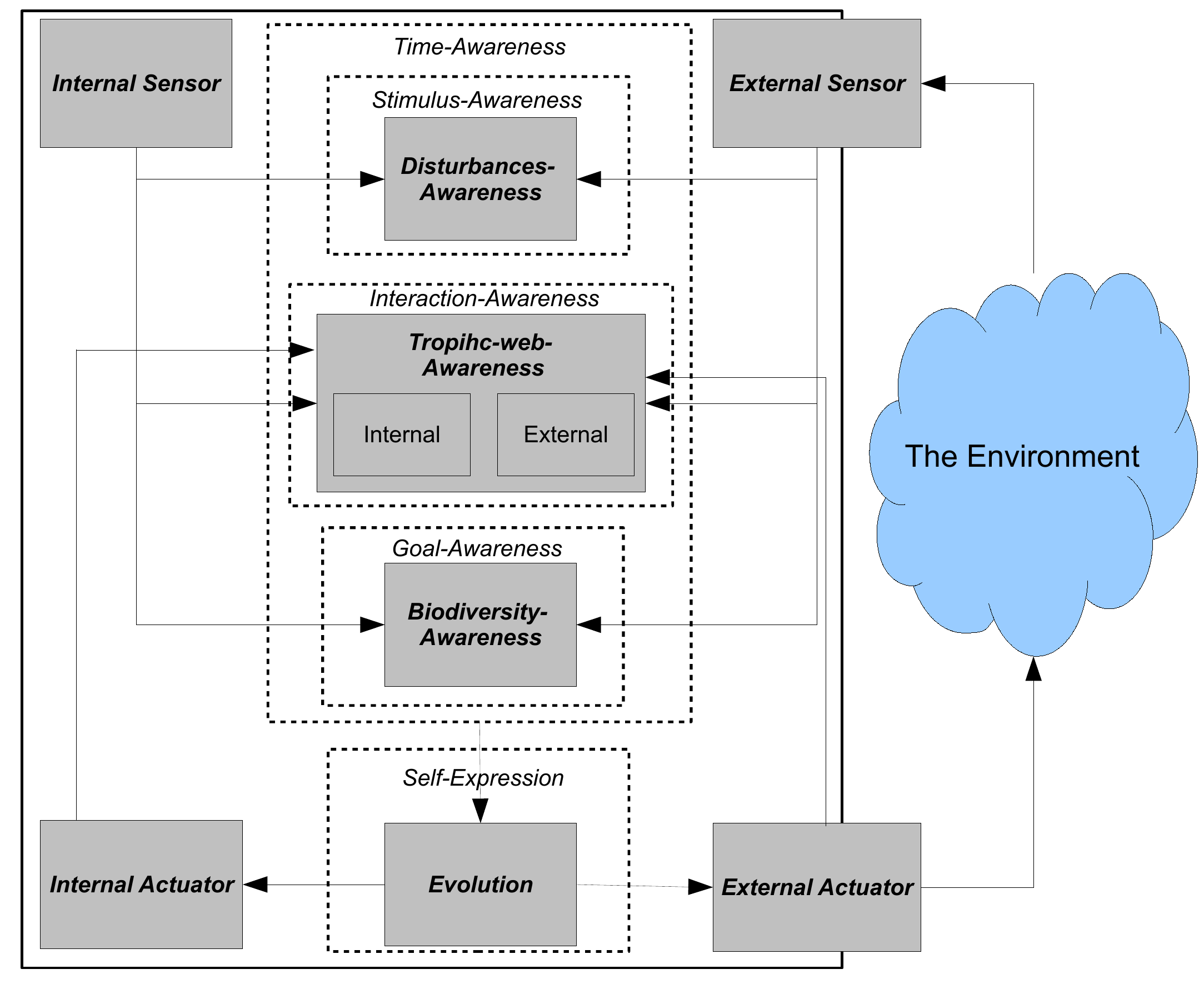}
\caption{The ecology-inspired pattern.}
\label{fig:pattern}
\end{figure}

\begin{itemize}

\item \textbf{\emph{Disturbances-Awareness:}} This is the basic level of awareness in an ecology-inspired pattern. It may acquire knowledge about either natural induced stress or human induced stress. It corresponds to the Stimulus-awareness in self-aware patterns.

\begin{itemize}
\item \emph{Example.} The cloud autoscaling process is able to sense different forms of disturbances, e.g., changes on workload, the neighbouring services/VMs, changes on pricing/requirements and availability of software/hardware resources. For instance, with disturbances-awareness, the cloud ecosystem is able to identify the source of stimulus and react upon. 
\end{itemize}

\item \textbf{\emph{Trophic-web-Awareness:}} Interactions within ecosystem are expressed in trophic web and therefore awareness of such a web is essential to capture the occurrences of possible interactions. In particular, this type of awareness could acquire knowledge about either internal or external interactions. For examples, the relationships between species in the corresponding group are clearly internal interactions; while those between different groups of species or even different ecosystems can be seen as external interactions. This awareness also permit co-evolution, i.e., two or more species might need to evolve in response to each others. For instance, co-evolution on two or more services implies that scaling decisions for one could have potential implications on the others. As a result, when making scaling decisions, implications on all services involved in the co-evolution needs to be catered for. To this end, Trophic-web-Awareness helps to measure, understand and quantify those implications, thus enabling the evolution of one service with respect to the others, leading to better informed decision making process. Trophic-web-Awareness corresponds to the Interaction-awareness in self-aware patterns.

\begin{itemize}
\item \emph{Example.} The cloud autoscaling process is able to aware of the relationships between cloud-based services and software/hardware resources, their functional dependency and the topology. Specifically, when disturbances are detected, cloud-based services would need to evolve in order to maintain stability. The knowledge of trophic web can help to explore the direction of evolution, e.g., assess the effects of certain way of evolution, and complain with the constraints imposed by dependency and topology.
\end{itemize}

\item \textbf{\emph{Biodiversity-Awareness:}} Increasing biodiversity is the key to improve stability and sustainability of an ecosystem. Awareness of biodiversity permits the ability to reason about and acknowledge the effects of evolutions w.r.t. stability and sustainability for the entire ecosystem. Therefore, it corresponds to the Goal-awareness in self-aware patterns.

\begin{itemize}
\item \emph{Example.} The cloud autoscaling process is able to reason about how different forms of evolution can affect the biodiversity which, in turn, influences the long-term stability and sustainability related to QoS, cost and energy consumption in the cloud ecosystem. Such awareness could produce explicit answers on \emph{when} the cloud should scale and \emph{what} is the amounts of scaling that lead to higher biodiversity. Concretely, biodiversity serves as the metric to assess the direction of evolution, thus providing guidance for the cloud ecosystem to maintain high biodiversity.
\end{itemize}

\item \textbf{\emph{Evolution:}} Evolution of an ecosystem can be regarded in two levels: (i) at micro evolution level, the species can evolve themselves to adapt to the environment or the other species. (ii) At macro evolution level, the species can change their habitats. It corresponds to the Self-expression in self-aware patterns.

\begin{itemize}
\item \emph{Example.} The cloud autoscaling process is able to know \emph{how} the cloud ecosystem can evolve. This is concerned with whether vertical scaling, horizontal scaling or both should be triggered.  Here, vertical scaling implies evolution at micro level, in which the cloud-based services can adapt their software configurations and hardware resource provisioning based on their demand. Similarly, horizontal scaling means evolution at macro level, e.g., adding/removing cloud-based services and VM migration/replication, which can cause changes to their habitats.
\end{itemize}

\end{itemize}

All the levels of awareness in cloud ecosystem can be connected to the Time-awareness from the self-aware patterns. This is because the natural ecosystem, which exists for millions of years, can often gain rich benefits from its long history. In the context of computing system, such historical data can be of relevant to the types of disturbances, changes of trophic web and evolutions of the biodiversity. As a result, historical knowledge plays an integral role in our ecology-inspired pattern, for example, evolutions of the biodiversity could gain insights from the past biodiversity levels, including their implication on the current cloud ecosystem and the entire path of evolution. This information can help to guide both the micro- and macro-level evolution.

\section {Opportunities and Challenges}

We have already shown that cloud ecosystem exhibits many similarities to the natural ecosystem. Presumably, when cloud autoscaling leads to higher level of biodiversity, the global stability and sustainability of cloud ecosystem would expected to be improved. This direction of research will create several opportunities and challenges:

\begin{itemize}

\item Autoscaling in the entire cloud is a complex and large-scale control problem. The notion of evolution from natural ecosystem can provide inspiration about how to ensure high biodiversity in the cloud ecosystem, and thus improve global stability and sustainability. However, incorporating the control mechanisms of biodiversity and cloud autoscaling is a research challenge. Additionally, selecting the right measurements and form of biodiversity for cloud ecosystem is also a difficult issue. We expect to obtain similar perception as our prior work~\cite{Chen:tsc:2016}\cite{Chen:2014:ucc} when applying ecological principles to this challenge.

\item Dynamics and uncertainty in the cloud significantly influence the design of an autoscaling system due to the time-varying workload, QoS interference and the behaviors of cloud-based services. We hope that the mechanisms and models form trophic web can better handle the dynamics and uncertainty. These mechanisms and models can provide us with insight about the interactions in different trophic levels and competition between species (i.e., cloud-based services). In addition, ecologists have applied several metrics (e.g., Shannon entropy) to quantify biodiversity in trophic web. The challenge is concerned with how the trophic web can be used to correlate the QoS with cloud configurations and resources; and how it might influence the decision making of autoscaling.

\item QoS interference and trade-offs are important issues in autoscaling decision making. Here, trade-offs do not only refer to the naturally conflicted objectives of the same cloud-based service (e.g., throughput vs. cost/energy), but also to the conflicted objectives for different cloud-based services caused by QoS interference. While QoS interference might be tackled using computational intelligence~\cite{Chen:seams:2014}\cite{Chen:2014}, it is still challenging to study how the insights of species competition and co-evolution can be used to resolve the trade-offs and to mitigate the effects of QoS interference in cloud. 

\end{itemize}

\section {Related Work}

There have been some successful attempts in applying ecological principles, theories and models to address issues in computer science. Examples can be found in the area of software engineering \cite{6747205}, collaborative adaptive systems \cite{6747203} \cite{mendez2013empirical}, distributed computing \cite{6335781} and grid computing \cite{gao2004novel}. However, there have been very limited efforts on adopting ecological view for cloud autoscaling.

Among others, ECOS \cite{6747205} is a research project that adopts ecological models to analyze the evolution of open-source software. Particularly, ecology-inspired methods are used to understand and better explain how the projects evolve, and what are the factors that drive the success of these projects. The goal is to optimize the fitness of software projects, leading to better stability of open-source software ecosystem.

Briscoe and Wilde \cite{6335781} intend to apply ecological thinking to create scalable and self-organizing approaches for distributed evolutionary computing. The aim is to maintain a stable evolution of the processes in distributed environment, i.e., what processes should run independently or incorrporately.

Ecology inspired approach has been used in grid computing \cite{gao2004novel}, with an aim to handle dynamics, uncertainty, diversity and evolution in grid services. However, they focus on subscription, discovery, selection and composition of grid-based services, as opposed to the cloud autoscaling of software configurations and hardware resources in our work.

The EU funded, multi-disciplinary project DEVERISTY \cite{6747203} is possibly the most related work to our research. DEVERISTY exploits the analogy with natural ecosystems in order to come to better principles and mechanisms for handling the emergence of diversity in collaborative adaptive systems. The objective is to increase diversity in software systems and thus achieving better stability, as well as the ability to react to unpredicted events \cite{mendez2013empirical}. Such an objective is consistent with our work, but we have particularly focused on the context of cloud computing and autoscaling, which exhibits some unique characteristics.

\section {Conclusion}

In this article, we have present an intuitive and sensible transposition of ecological perspective to the context of cloud autoscaling. Deriving from existing self-aware software patterns, we have also proposed an architecture pattern for enabling such a transition, with respect to the different principles of natural ecosystem, including disturbance, Trophic-web, biodiversity and (co-) evolution. The challenges for this direction of research are also discussed.

Cloud computing will continue to attract more and more participates for its scalable, elastic and on-demand promises. Stability and sustainability can quickly become critical quality indicators of cloud-based services, leading to several new challenges. Until recently, there has been an increasing interest in investigating the assurance of long term benefits in cloud via autoscaling. To the best of our knowledge, there has been no known method that is explicitly developed for the long term benefit of stability and sustainability of cloud-based services. In this premise, the well-established ecological principles appear to be neat solution for those challenges, as we have discussed in this article. This will further advance the existing view of cloud and its autoscaling.

As future work, we aim to explicitly tackle the aforementioned research challenges, particularly focusing on the design of ecology driven mechanisms to handle dynamics and uncertainty exhibited in the cloud.

\section {Acknowledgment}
This work is supported by the Paul and Yuanbi Ramsay Research Funding Award form the School of Computer Science, University of Birmingham, UK.


\bibliography{references}
\bibliographystyle{abbrv}
\nocite{*}
\end{document}